\documentclass[12pt]{article}
\usepackage{amssymb}
\usepackage[dvips]{epsfig}

\newcommand{\be}{\begin{equation}}
\newcommand{\ee}{\end{equation}}
\def\bea{\begin{eqnarray}}
\def\eea{\end{eqnarray}}

\newcommand{\bn}{\begin{eqnarray}}
\newcommand{\en}{\end{eqnarray}}

\newcommand{\p}{\partial}

\newcommand{\nn}{\nonumber}

\newcommand{\no}{\noindent}

\newcommand{\s}{\,\,\,\,}

\def\bea{\begin{eqnarray}}
\def\eea{\end{eqnarray}}

\newcommand{\beq}{\begin{eqnarray}}
\newcommand{\eeq}{\end{eqnarray}}
\begin{document}

\title{\textbf{Master actions for massive spin-3 particles in D=2+1}}
\author{E.L. Mendon\c ca\footnote{eliasleite@feg.unesp.com}, D. Dalmazi\footnote{dalmazi@feg.unesp.br},
\\
\textit{{UNESP - Campus de Guaratinguet\'a - DFQ} }\\
\textit{{Avenida Dr. Ariberto Pereira da Cunha, 333} }\\
\textit{{CEP 12516-410 - Guaratinguet\'a - SP - Brazil.} }\\}
\date{\today}
\maketitle

\begin{abstract}

 We present here a relationship among massive self-dual models for spin-3 particles in $D=2+1$ via the master action procedure. Starting with a first order model (in derivatives) $S_{SD(1)}$ we have constructed a master action which interpolates among a sequence of four self-dual models $S_{SD(i)}$ where $i=1,2,3,4$. By analyzing the particle content of mixing terms, we give additional arguments that explain why it is apparently impossible to jump from the fourth order model to a higher order model. We have also analyzed similarities and differences between the fourth order $K$-term in the spin-2 case and the analogous fourth order term in the spin-3 context.

\end{abstract}

\newpage

\section{ Introduction}
Higher spin massive particles are present in the spectrum of string theories.
Although we have not observed higher spin elementary particles in nature yet,
resonant states have been detected, see \cite{cern} as a recent example of
spin-3 resonance observation. In general, Lagrangians for massive particles in
$D=4$ have the same form in arbitrary dimensions, like e.g. the Maxwell-Proca
(spin-1) and Fierz-Pauli (spin-2) theories and even higher spin theories. Since
in $D=3$ one can build up such Lagrangians, see \cite{banerjee,solda1}, for the
spin-1 and spin-2 cases respectively, via soldering of opposite helicity
states, one may claim that the Lagrangians which represent helicity eigenstates
in $D=3$ are the basic building blocks for bosonic massive higher spin
particles Lagrangians. Thus, the study of such parity breaking theories
(self-dual models) may have some interesting connection with massive higher
spin particles in arbitrary dimensions. Here we investigate the simplest case
after the spin-2 one, i.e., the massive spin-3 states in $D=3$. Based on our
previous experience with massive spin-2 gauge theories
\cite{dualdescription,newself}, we are going to analyze it  by means of the
master action technique.

Recently, we have addressed \cite{NGEs3} this subject through another dualization procedure, the Noether Gauge
Embedmet NGE, which is based on the existence of a local symmetry in the highest derivative term of the
self-dual model which is not present in the lower derivative terms. In complete analogy with the spin-2 case we
have shown that starting with the first order non gauge invariant self-dual model \cite{AragoneS31} it is
possible to obtain the second \cite{AragoneS32}, third \cite{deserdam} and a fourth order self-dual models,
where the last one has all the auxiliary fields needed to correctly describe only one helicity $+3$ or $-3$
particle. In \cite{NGEs3} we have faced the problem of missing gauge symmetries which are required in order to
proceed with the technique and go beyond the fourth order self-dual model, which might be naively expected since
looking at the spin-1 and spin-2 examples, one can see that there are two and four self-dual descriptions for
the singlets respectively, indicating that there might be some $2s$ rule for the highest order of the spin-$s$
self-dual model, where $s$ is the spin. In fact the authors in \cite{BHThs} have proposed fifth and sixth order
equations of motion, for a massive spin-3 particle, however without considering auxiliary fields, where however
the fifth-order model contains ghosts.

Here, we come back to this point. We find an alternative explanation on why it is apparently impossible to
complete the chain of $2s$ models in the spin-3 case, and give a demonstration that the classical equivalence
among first, second, third and fourth order spin-3 models obtained in \cite{NGEs3}, holds also at the quantum
level.

In the master action approach a fundamental ingredient consists of finding appropriate mixing terms between the
dual fields, which must be free of particle content. So, here we organize the paper first providing a discussion
on the particle content of mixing terms for spins two and three. We then, propose a master action that
interpolates among the first three spin-3 self-dual models, and obtain their dual maps. As a last step we show
that is possible to construct a new master action using only totally symmetric fields, which interpolates
between the third and the fourth order self-dual models. In the last section final remarks on particle content
of the fourth order term clarity the difficulties in going beyond the fourth order self-dual model.

\section{Trivial Lagrangians}
\subsection{Chain of self-dual models}

In order to capture the essential features of the master action \cite{DJ} approach used here in $D=2+1$ to jump
from a $k$-th order self-dual model ${\cal L}_{SD}^{(k)}$ to ${\cal L}_{SD}^{(k+1)}$, we have found instructive
to write down a toy model version of ${\cal L}_{SD}^{(k)}$ in a symbolic notation suppressing all Lorentz
indices and complicated details about the different operators appearing in the Lagrangian densities, namely,

\be {\cal L}_{SD}^{(k)}[A]=m^2b_k\left[(-1)^k\, A\, \hat{\p}^k A+ A\, \hat{\p}^{k-1}A\right].\label{1}\ee

\no For convenience we define a dimensionless derivative $\hat{\p}=\p/m$ and $b_k=(-1)^{k(k+1)/2}$ is such that
it satisfies:

\be b_{k+1}=(-1)^{k+1}b_k.\label{6}\ee

\no The first order theory ${\cal L}_{SD}^{(1)}[A]$ may represent the self-dual model of \cite{ad1} in the
spin-1 case or the self-dual models of \cite{AragoneS31} and \cite{AragoneS21} in the spin-2 and spin-3 cases
respectively, while ${\cal L}_{SD}^{(2)}$  represents the spin-1 Maxwell-Chern-Simons ($MCS$) model of
\cite{DJT} as well as the second order models of spin-2 and spin-3 defined in \cite{desermac} and
\cite{AragoneS32} respectively. The lagrangian density  ${\cal L}_{SD}^{(3)}$  stands for both the linearized
topologically massive gravity of \cite{DJT} and the third order spin-3 model of \cite{deserdam}. Finally  ${\cal
L}_{SD}^{(4)}$ may  represent either the linearized version of the higher derivative topologically massive
gravity of \cite{newself, Tow2} or the spin-3 fourth-order model whose main equations of motion are given in
\cite{BHThs} and the whole action in \cite{NGEs3}. More precisely, the $A$-field in (\ref{1}) stands for a
rank-$s$ tensor field and in the spin-s case while ${\cal L}_{SD}^{(k)}$ corresponds only to the two terms of the
self-dual model which are quadratic in the rank-$s$ field.

The basic idea of the master action approach is to add to ${\cal L}_{SD}^{(k)}$ a ``mixing term'' between the
$A$-field and the dual $B$-field and define a master model:

\be {\cal L}_{M}[A,B]=m^2b_k\left[(-1)^k\, A\, \hat{\p}^k A+ A\, \hat{\p}^{k-1}A-(-1)^k\, (A-B)\, \hat{\p}^k
(A-B)\right].\label{2}\ee

\no The mixing term is essentially the highest derivative term of ${\cal L}_{SD}^{(k)}$. After the trivial shift
$B\to \tilde{B}+A$, which produces a trivial Jacobian in the path integral, we have:

\be {\cal L}_{M}={\cal L}_{SD}^{(k)}[A]+ m^2b_k(-1)^{k+1}\, \tilde{B}\, \hat{\p}^k \tilde{B}.\label{3}\ee

\no On the other hand, we can rewrite (\ref{2}) neglecting total derivatives:

\be {\cal L}_{M}[A,B]=m^2b_k\left\lbrace B \,\hat{\p}^{k+1}\,B- (-1)^kB\,
\hat{\p}^k\,B+\left[A-(-1)^k\hat{\p}B\right]\hat{\p}^{k-1}\left[A-(-1)^k\hat{\p}B\right]\right\rbrace.\label{5}\ee

After the shift $A\to \tilde{A}+(-1)^k\hat{\p}B$ and use of (\ref{6}) we have:

\be {\cal L}_{M}={\cal L}_{SD}^{k+1}[B]+m^2b_k\tilde{A}\,\hat{\p}^{k-1}\tilde{A}\label{7}\ee

Therefore if both $\tilde{B}\,\hat{\p}^{k}\,\tilde{B}$ and $\tilde{A}\,\hat{\p}^{k-1}\,\tilde{A}$ have no particle
content, it is clear that (\ref{3}) and (\ref{7}) establish the physical equivalence (duality) of ${\cal
L}_{SD}^{(k)}$ and ${\cal L}_{SD}^{(k+1)}$. It amounts to assume that ${\cal L}_{SD}^{(k)}$ in (\ref{1}) is made
out of two terms without physical content (trivial). In the next subsection we review  the triviality of some
spin-2 lagrangian densities as an introduction to the spin-3 case of the subsection 2.3.

\subsection{spin-2 and spin-3}
\subparagraph{} 
As a warm up let us briefly review the very known terms that can be used as mixing terms in the spin-2  context. First we have a Chern-Simons like term of first order in derivatives:

\be S_{CS2}^{(1)}=\frac{m}{2}\int d^3x\,
\epsilon^{\mu\nu\alpha}f_{\mu\beta}\,\p_{\nu}\,f_{\alpha}^{\s\beta},\label{tcs2}\ee

\no where $f_{\mu\nu}$ is a non-symmetric tensor. Similar to the spin-1 case one can show that the general
solution to the equations of motion is also pure gauge, with $f_{\mu\nu}=\p_{\mu}\xi_{\nu}$.

We can also use the linearized version of the Einstein-Hilbert term, which is second order in derivatives. We
write it in the following way:

\be S_{EH}=-\frac{1}{2}\int d^3x\, f_{(\mu\nu)}E^{\mu\alpha}E^{\nu\beta}f_{(\alpha\beta)},\label{rl2}\ee

\no from the equations of motion with respect to $f_{(\mu\nu)}$ \footnote{Where
$f_{(\mu\nu)}=(f_{\mu\nu}+f_{\nu\mu})/2$.} we have the second order differential equations $
E^{\mu\alpha}E^{\nu\beta}f_{(\alpha\beta)}=0$. By applying twice the
Levi-Civita symbol in  the equations of motion, we have:

\be
\epsilon_{\mu\gamma\rho}\epsilon_{\nu\lambda\xi}E^{\mu\alpha}E^{\nu\beta}f_{(\alpha\beta)}=R^{L}_{\gamma\rho\lambda\xi}(f)=0.\label{rl1}\ee

\no Where $R^{L}_{\gamma\rho\lambda\xi}(f)$ stands for the linearized Riemann tensor. The general solution of
(\ref{rl1}) (see \cite{DJT}) is also pure gauge $f_{(\mu\nu)}=\p_{\mu}\Phi_{\nu}+\p_{\nu}\Phi_{\mu}$.

Finally, besides (\ref{tcs2}) and (\ref{rl2}) one has yet a third option, which is the third order
topological Chern-Simons term whose linearized version is: \be S_{CS2}^{(3)}=\frac{1}{2m}\int
d^3x\,f^{(\mu\nu)}\Box\theta_{\mu\alpha}E_{\nu\beta}f^{(\alpha\beta)}.\ee

\no Where we have introduced the transverse projection operator
$\theta_{\mu\nu}=(\eta_{\mu\nu}-\p_{\mu}\p_{\nu}/\Box)$. In \cite{DJT} the authors demonstrate through the
helicity decomposition method that this term has no particle content. So, it is possible to use it as a mixing
term.

For spin-3 the first candidate for a mixing term is  the first-order Chern-Simons like term, which was first
introduced in \cite{AragoneS31}:

\be S_{CS3}^{(1)}=\frac{m}{2}\int
d^3x\,\epsilon_{\mu\nu\alpha}\omega^{\mu(\beta\gamma)}\p^{\nu}\omega^{\alpha(\beta\gamma)},\label{cs31}\ee

\no where  $\omega_{\mu(\beta\gamma)}$, is symmetric and traceless i.e
$\omega_{\mu(\beta\gamma)}=\omega_{\mu(\gamma\beta)}$ and $\eta^{\beta\gamma}\omega_{\mu(\beta\gamma)}=0$ with
$\omega_{\gamma}\equiv\eta^{\mu\beta}\omega_{\mu(\beta\gamma)}$. For the same reasons mentioned before it is not
difficult to convince oneself that the general solution of the equation of motion derived from (\ref{cs31}) is
pure gauge, $\omega_{\mu(\beta\gamma)}=\p_{\mu}\Lambda_{(\beta\gamma)}$ where
$\eta^{\beta\gamma}\Lambda_{(\beta\gamma)}=0$.

We can also take the usual Singh-Hagen second order massless spin-3 term \cite{Fronsdal} as a mixing term, it
comes into the game as an analogue of the spin-2 Einstein-Hilbert term. Just as in the case of spin-2, one can
check that it is possible to write it in terms of partially symmetric tensors or in terms of totally symmetric
tensors:

\be S_{SH}=\frac{1}{2}\int d^3x\,\,\xi_{\mu(\beta\gamma)}\Omega^{\mu(\beta\gamma)}(\xi)=\frac{1}{2}\int d^3x\,\,
\phi_{\mu\nu\alpha}G^{\mu\nu\alpha}(\phi),\label{SH}\ee

\no  where:

\be \Omega_{\mu(\beta\gamma)}(\xi)=
3(\xi_{\beta(\mu\gamma)}+\xi_{\gamma(\mu\beta)}-\xi_{\mu(\beta\gamma)})-2\eta_{\beta\gamma}\xi_{\mu}\quad,\quad
\xi_{\mu(\beta\gamma)}=E_{\mu}^{\s\lambda}\omega_{\lambda(\beta\gamma)}.\ee


On the right hand site of (\ref{SH}) we
have used the spin-3 ``Einstein tensor'' given in terms of the totally symmetric field $\phi_{\mu\beta\gamma}$
and defined in \cite{dWF,deserdam} as:

\be G^{\mu\beta\gamma}(\phi)\equiv R^{\mu\nu\lambda}-\frac{1}{2}\eta^{(\mu\nu}R^{\lambda)},\label{Gtensor}\ee

\no where the ``Ricci'' tensor is given by  $R^{\mu\nu\lambda}=\Box
\phi^{\mu\nu\lambda}-\p_{\alpha}\p^{(\mu}\phi^{\alpha\nu\lambda)}+\p^{(\mu}\p^{\nu}\phi^{\lambda)}$ and its
trace $R^{\lambda}=\eta_{\mu\nu}R^{\mu\nu\lambda}$.

Finally, the equivalence between the two notations in (\ref{SH}) is possible thanks to the general decomposition
bellow:

\be \omega_{\mu(\beta\gamma)}=
\frac{1}{\sqrt{3}}\left[\phi_{\mu\beta\gamma}+\frac{1}{4}(\eta_{\lambda\beta}\phi_{\gamma}+\eta_{\lambda\gamma}\phi_{\beta})-\frac{1}{2}\eta_{\beta\gamma}\phi_{\lambda}\right]+
(\epsilon_{\mu\nu\beta}\chi^{\nu}_{\s\gamma}+\epsilon_{\mu\nu\gamma}\chi^{\nu}_{\s\beta}),\label{deco}\ee

\no where  $\chi_{\mu\nu}(x)=\chi_{\nu\mu}(x)$ and $\eta^{\mu\nu}\chi_{\mu\nu}=\chi=0$. The numerical factors in (\ref{deco}) are obtained in such a way that our
results fit the results of  \cite{deserdam}. Besides, one can verify that both sides of (\ref{deco}) have the
same number of independent components in $D=2+1$.

From the equations of motion derived from (\ref{SH}) with respect to $\phi_{\mu\beta\gamma}$ we conclude that
the Einstein tensor (\ref{Gtensor}) vanishes:

\be G_{\mu\nu\lambda}(\phi)=0,\label{g=0}\ee

\no which implies that the ``Ricci'' tensor is null $R_{\mu\nu\lambda}=0$. In \cite{deserdam} the authors
demonstrate that, as in the spin-2 case, in $D=2+1$ the curvature Riemann tensor for spin-3
\bea R_{\mu\nu\lambda\alpha\beta\gamma}(\phi)&\equiv&\p_{\alpha}\p_{\beta}\p_{\gamma}\phi_{\mu\nu\lambda}-\p_{\mu}\p_{\beta}\p_{\gamma}\phi_{\alpha\nu\lambda}-\p_{\alpha}\p_{\beta}\p_{\lambda}\phi_{\mu\nu\gamma}+\p_{\mu}\p_{\beta}\p_{\lambda}\phi_{\alpha\nu\gamma}\nn\\
&-&\p_{\alpha}\p_{\nu}\p_{\gamma}\phi_{\mu\beta\lambda}+\p_{\mu}\p_{\nu}\p_{\gamma}\phi_{\alpha\beta\lambda}+\p_{\alpha}\p_{\nu}\p_{\lambda}\phi_{\mu\beta\gamma}-\p_{\mu}\p_{\nu}\p_{\lambda}\phi_{\alpha\beta\gamma},\label{riemanns3}\eea
\no can be expressed in terms of the Ricci tensor since the Weyl tensor vanishes in $D=2+1$, thus
$R_{\mu\nu\lambda\alpha\beta\gamma}=0$ follows from $R_{\mu\nu\lambda}=0$. The general solution for the null
curvature Riemann tensor is pure gauge $\phi_{\mu\nu\lambda}=\p_{(\mu}\tilde{\Lambda}_{\beta\gamma)}$ where
$\tilde{\Lambda}=\eta^{\mu\nu}\tilde{\Lambda}_{\mu\nu}=0$.  Note that $\delta_{\xi}S_{SH}=0$ where
$\delta_{\xi}\omega_{\mu(\beta\gamma)}=\p_{\mu}\xi_{(\beta\gamma)}$ with
$\eta^{\beta\gamma}\tilde{\xi}_{(\beta\gamma)}=0$. Then, the action (\ref{SH}) has no particle content in
$D=2+1$.

Besides the first two terms (\ref{cs31}) and (\ref{SH}) introduced before, one can also use as a mixing term the
third order spin-3 Chern-Simons term, which can be written  in terms of a partially symmetric tensor or a totally
symmetric tensor:

\bea S_{CS3}^{(3)}&=&\frac{1}{2m}\int d^3x\,
\Omega_{\mu(\beta\gamma)}(\xi)E^{\mu}_{\s\lambda}\Omega^{\lambda(\beta\gamma)}(\xi)=\frac{3}{2m}\int
d^3x\,\phi_{\mu\nu\lambda}E^{\mu}_{\s\gamma}G^{\gamma\nu\lambda}(\phi)\nn\\
&=&\frac{1}{2m}\int d^3x\, C_{\mu\beta\gamma}(\phi)G^{\mu\beta\gamma}(\phi).\label{action}\eea

 \no We have used the symmetrized curl defined in \cite{deserdam}, given by:
 
\be C_{\mu\beta\gamma}(\phi)\equiv E_{\mu}^{\s\s\nu}\phi_{\nu\beta\gamma}+E_{\beta}^{\s\s\nu}\phi_{\nu\beta\mu}+E_{\gamma}^{\s\s\nu}\phi_{\nu\beta\mu}\ee

and $G^{\mu\beta\gamma}$ is given in
(\ref{Gtensor}). The authors of \cite{BHThs} have suggested that the more natural analogue of the Einstein
tensor for spin-3 should be a rank-3 third order tensor instead of (\ref{Gtensor}). This alternative is
particularly useful for us, since defining $ {\cal
G}_{\mu\nu\rho}\equiv E_{\mu}^{\s\alpha}E_{\nu}^{\s\beta}E_{\rho}^{\s\gamma}\phi_{\alpha\beta\gamma}$ the action
(\ref{action}) becomes: \be S_{CS3}^{(3)}=\frac{1}{2m}\int d^3x\,  \phi_{\mu\nu\lambda}{\cal
G}^{\mu\nu\lambda}=\frac{1}{2m}\int d^3x\,
\phi_{\mu\nu\lambda}E^{\mu\alpha}E^{\nu\beta}E^{\lambda\gamma}\phi_{\alpha\beta\gamma},\label{towcs}\ee

\no and makes evident the gauge symmetry $\delta \phi_{\alpha\beta\gamma}=\p_{(\alpha}\Lambda_{\beta\gamma)}$
with $\Lambda_{\beta\gamma}=\Lambda_{\gamma\beta}$ an arbitrary symmetric parameter, compare (\ref{rl2}) with
(\ref{towcs}). Finally by taking the equations of motion from (\ref{towcs}) ${\cal G}_{\mu\nu\lambda}=0$, we
have the pure gauge solution $\phi_{\mu\nu\lambda}=\p_{(\mu}\Lambda_{\nu\lambda)}$. One can notice that like in
the spin-2 case, once we have:

\be \epsilon_{\gamma\mu\nu}\epsilon_{\delta\lambda\alpha}\epsilon_{\rho\beta\gamma}{\cal
G}^{\gamma\delta\rho}=0,\ee

\no these results in $ R_{\mu\nu\lambda\alpha\beta\gamma}=0$. Then, again according to the theorem demonstrated
in \cite{deserdam} one can verify that the third order Chern-Simons term (\ref{towcs}) has no particle content,
so it can be used as a mixing term in order to construct a master action just like (\ref{cs31}) and (\ref{SH}).

\section{First, second and third order spin-3 self-dual models}
\subparagraph{}

The master action is constructed from the first order self-dual model proposed in \cite{AragoneS31} which is the
spin $3$ analogue of the spin $2$ and spin $1$ self-dual models of \cite{AragoneS21,ad1} respectively:

\be S_{SD(1)}[\omega, A]=\int d^3x\,\, \left\lbrack
\frac{m}{2}\epsilon^{\mu\nu\alpha}\omega_{\mu(\beta\gamma)}\p_{\nu}\omega_{\alpha}^{\s(\beta\gamma)}+\frac{m^2}{6}(\omega_{\mu}\omega^{\mu}-\omega_{\mu(\beta\gamma)}\omega^{\beta(\mu\gamma)})
+ m^2 \omega_{\mu}A^{\mu}\right\rbrack+S_1[A],\label{sd1}\ee

\no where \be S_1[A]=\int
d^3x\,\,\left\lbrack-9m\,\,\epsilon^{\mu\nu\alpha}A_{\mu}\p_{\nu}A_{\alpha}-9m^2\,\,A_{\mu}A^{\mu}
\,-12(\p_{\mu}A^{\mu})^2 \right\rbrack,\label{lagaux1}\ee

\no is the required auxiliary action such that (\ref{sd1}) describes only one massive spin-3 particle. The
Fierz-Pauli conditions can be obtained from the equations of motion derived from (\ref{sd1}), demonstrations can
be found in \cite{NGEs3,AragoneS31}.

By adding mixing terms without particle content we aim to construct a master action from (\ref{sd1}). We can use
the first order Chern-Simons term (\ref{cs31}), the usual massless second order spin-3 term (\ref{SH}) and the
third order Chern-Simons term (\ref{action}) as mixing terms. However as we have observed in \cite{NGEs3} when
we get the third order self-dual model the whole action can be described by totally symmetric tensors through
the decomposition (\ref{deco}), so, first we are going to construct a master action interpolating among the first
three self-dual models, and then as a last step an action interpolating between the third-order self-dual model
and the fourth order self-dual model both of them in terms of totally symmetric tensors. The first master action
is suggested as follows:

\bea S_M&=& \int \left[\frac{m}{2}\omega\cdot d\omega +\frac{m^2}{6}(\omega^2)-\frac{m}{2}(\omega-g)\cdot
d(\omega-g) + \frac{1}{2} (h-g)\cdot d \Omega(h-g)\right]\nn\\ &+& m^2\int d^3x\,\,\omega_{\mu}A^{\mu}+
S_1[A],\label{mestra}\eea

\no where $g_{\mu(\beta\gamma)}$ and $h_{\mu(\beta\gamma)}$ are new auxiliary fields. Here, we use the same
shorthand notation adopted in \cite{dualdescription} where:
\bea \int \,\, (\omega^2)&\equiv& \int d^3x\,\, (\omega_{\mu}\omega^{\mu}-\omega_{\mu(\beta\gamma)}\omega^{\beta(\mu\gamma)}),\\
 \int \,\, \omega\cdot d\omega&\equiv&\int d^3x\,\,\epsilon^{\mu\nu\alpha}\omega_{\mu(\beta\gamma)}\p_{\nu}\omega_{\alpha}^{\s(\beta\gamma)},\\
 \int\,\, \omega\cdot d\Omega(\omega)&\equiv&\int d^3x\,\,\xi_{\mu(\beta\gamma)}(\omega)\,\Omega^{\mu(\beta\gamma)}\,\left\lbrack\xi(\omega)\right\rbrack.\eea

In order to interpolate among the dual models, obtaining dual maps let us introduce a source term
$j_{\mu(\beta\gamma)}$ and define the generating functional:

\be W_M[j]=\int {\cal D}\omega \,{\cal D} g \,{\cal D} h \,{\cal D} A \, \exp \, i\left(S_M+\int d^3x\,
j_{\mu(\beta\gamma)}\omega^{\mu(\beta\gamma)}\right).\ee \no The first thing we note is that in order to recover
the first-order self-dual model we just need to make the shifts $h\to h+g$ and $g\to g+\omega$ in
(\ref{mestra}). Then we get the mixing terms decoupled. Since they have no particle content, see (\ref{cs31})
and (\ref{SH}), we end up with the content of the first order self-dual model (\ref{sd1}). So deriving with
respect to the sources we find the following identity: \be \langle
\omega_{\mu_1(\beta_1\gamma_1)}(x_1)\,\,...\,\,\omega_{\mu_N(\beta_N\gamma_N)}(x_N)\rangle_M= \langle
\omega_{\mu_1(\beta_1\gamma_1)}(x_1)\,\,...\,\,\omega_{\mu_N(\beta_N\gamma_N)}(x_N)\rangle_{SD(1)}.\ee

On the other hand making only the shift $h \to h+g$ and then functionally integrating over $h$ we have:


\be  S_M=\int
d^3x\,\, \omega_{\mu(\beta\gamma)} \tilde{g}^{\mu(\beta\gamma)} +\int \left(
\frac{m^2}{6}(\omega^2)-\frac{m}{2}g\cdot d g\right)+S_1[A], \label{mestra3} \ee

\no where: \be
\tilde{g}^{\mu(\beta\gamma)}=m\epsilon^{\mu\nu\alpha}\p_{\nu}g_{\alpha}^{\s(\beta\gamma)}+\frac{m^2}{2}f^{\mu(\beta\gamma)}(A)+j^{\mu(\beta\gamma)},\label{gtio}\ee

\no with:

\be
f^{\mu(\beta\gamma)}(A)=\eta^{\beta\mu}A^{\gamma}+\eta^{\gamma\mu}A^{\beta}-\frac{2}{3}\eta^{\beta\gamma}A^{\mu}.\label{fa}\ee

\no In  (\ref{mestra3}) we have a quadratic term and a linear term in $\omega$. This suggests an
integration over  $\omega$ in such a way that we obtain an action for $g$.

Due to the absence of particle content of terms like (\ref{cs31}) we have a second order self-dual action: \be S_{SD(2)}= \int
\left\lbrack-\frac{1}{2}g\cdot d\Omega(g)-\frac{m}{2}g\cdot dg -
 \frac{m}{2} f(A)\cdot d g +j_{\mu(\beta\gamma)}F^{\mu(\beta\gamma)}(g,A)+{\cal{O}}(j^2)\right\rbrack
+S_2[A],\label{mestra5}\ee

\no where ${\cal O}(j^2)$ stands for quadratic terms in the source and: \be
F^{\mu(\beta\gamma)}(g,A)=\frac{\Omega^{\mu(\beta\gamma)}\left\lbrack\xi(g)\right\rbrack}{m}+f^{\mu(\beta\gamma)}(A).\ee

The action (\ref{mestra5}) is exactly the one for the second order self-dual model given in \cite{AragoneS32}, except
for the source term it automatically includes, the new action for the auxiliary field   $A_{\mu}$ given by: \be
S_2[A]=\int
d^3x\,\,\left\lbrack-9m\,\,\epsilon^{\mu\nu\alpha}A_{\mu}\p_{\nu}A_{\alpha}-\frac{32m^2}{3}\,\,A_{\mu}A^{\mu}
\,-12(\p_{\mu}A^{\mu})^2 \right\rbrack.\label{lagaux2}\ee Now, deriving with respect to the source in
(\ref{mestra}) and (\ref{mestra5}) we have the correlation functions duality: \be \langle
\omega_{\mu_1(\beta_1\gamma_1)}(x_1)\,\,...\,\,\omega_{\mu_N(\beta_N\gamma_N)}(x_N)\rangle_{SD1}= \langle
F_{\mu_1(\beta_1\gamma_1)}(x_1)\,\,...\,\,F_{\mu_N(\beta_N\gamma_N)}(x_N)\rangle_{SD2} +\,\, C.T,\label{map1}
\ee

\no where $C.T$ stands for contact terms. The relation (\ref{map1}) gives us the dual map at classical and
quantum level: \be \omega^{\mu(\beta\gamma)}\longleftrightarrow F^{\mu(\beta\gamma)}(g,A).\label{dualmap1}\ee

\no Moreover, one can easily demonstrate that the interaction term between  the spin-3 field and the vector
field is the same one obtained previously through the NGE procedure in \cite{NGEs3} i.e.,
$-\frac{m}{2}f(A)\cdot dg=2m\xi_{\mu}(g)A^{\mu}$. The equations of motion of the first order self-dual model
$S_{SD(1)}$ can be dual mapped into the equations of motion for the second order self-dual model $S_{SD(2)}$ via (\ref{dualmap1}) as
we have demonstrated in \cite{NGEs3}.

Considering again the master action written in (\ref{mestra}), without shifting $h\to h+g$, instead of
(\ref{mestra5}), the master action would be:
\bea S_M&=&\int \left\lbrack-\frac{1}{2}g\cdot d\Omega(g)-\frac{m}{2}g\cdot dg+\frac{1}{2}(h-g)\cdot d\Omega(h-g)\right.\nn\\
&-& \left. \frac{m}{2} f(A)\cdot dg -j_{\mu(\beta\gamma)}F^{\mu(\beta\gamma)}(g,A)+{\cal{O}}(j^2)\right\rbrack
+S_2[A]. \label{mestra61}\eea

\no It is straightforward to show that: \be\int\,h\cdot d\Omega(g)=\int g\cdot d\Omega(h).\ee


\no So, we can rewrite (\ref{mestra61}) in such a way that:
\bea S_M&=& \int \,\left\lbrack -\frac{m}{2}(g+C)\cdot d(g+C)+\frac{m}{2}C\cdot dC+\frac{1}{2}h\cdot d\Omega(h)-j_{\mu(\beta\gamma)}F^{\mu(\beta\gamma)}(g,A)+{\cal{O}}(j^2)\right\rbrack\nn\\
&+& S_2[A],\label{mestra6}\eea

\no where we have defined: \be C=\frac{\Omega(h)}{m}-\frac{\Omega(j)}{m^2}+f(A).\ee

\no The shifts $g\to g-C$ and $g_{\mu(\beta\gamma)}\to
3(j_{\beta(\mu\gamma)}+j_{\gamma(\mu\beta)}-j_{\mu(\beta\gamma)})-2\eta_{\beta\gamma}j_{\mu}$ in (\ref{mestra6})
will completely decouple $g$ from $C$ and $j$. Then we can integrate over $g$. Substituting back $C$ we have the
third order self-dual action of \cite{deserdam}:

\bea S_{SD(3)}&=&\int \,\left[\frac{1}{2}h\cdot d\Omega(h)+\frac{1}{2m}\Omega(h)\cdot d\Omega(h)+f(A)\cdot
d\Omega(h)
+ j_{\mu(\beta\gamma)}H^{\mu(\beta\gamma)}(h,A)+{\cal{O}}(j^2)\right]\nn\\
&+& S_3[A],\label{mestrafinal}\eea

\no where $H^{\mu(\beta\gamma)}(h,A)$ give us the dual map:

\be \omega^{\mu(\beta\gamma)}\longleftrightarrow
H^{\mu(\beta\gamma)}\ee

\no where:
\be H^{\mu(\beta\gamma)}\equiv -\frac{1}{m}\Omega^{\mu(\beta\gamma)}\left[\frac{\Omega[\xi(h)]}{m}+f(A)\right]+f^{\mu(\beta\gamma)}(A).\label{dual3}\ee

 \no Again, the auxiliary action is automatically redefined and given in agreement with \cite{deserdam} by:
 \be S_3[A]=\int d^3x\,\,\left\lbrack-\frac{32 m}{3}\,\,\epsilon^{\mu\nu\alpha}A_{\mu}\p_{\nu}A_{\alpha}-\frac{32m^2}{3}\,\,A_{\mu}A^{\mu}
 \,-12(\p_{\mu}A^{\mu})^2 \right\rbrack.\label{lagaux3}\ee

\no By deriving with respect to the source term in (\ref{mestra}) and (\ref{mestrafinal}) we have the
equivalence of the correlation functions: \be \langle
\omega_{\mu_1(\beta_1\gamma_1)}(x_1)\,\,...\,\,\omega_{\mu_N(\beta_N\gamma_N)}(x_N)\rangle_{SD1}= \langle
H_{\mu_1(\beta_1\gamma_1)}(x_1)\,\,...\,\,H_{\mu_N(\beta_N\gamma_N)}(x_N)\rangle_{SD3} + C.T.\label{map2}\ee

In the next section we are going to perform the interpolation between the third order self-dual model and the
new fourth order self-dual model.

\section{Master action in terms of totally symmetric fields}
From now on, we propose a new master action only in terms of totally symmetric fields. It can be constructed
from the third order self-dual model obtained before, by means of the decomposition (\ref{deco}) in (\ref{mestrafinal}), with $\omega_{\mu(\beta\gamma)}$ replaced by $h_{\mu(\beta\gamma)}$,

 \bea S_{SD(3)}[\phi,A]&=&\int d^3x\left\lbrack
-\frac{1}{2}\phi_{\mu\beta\gamma} G^{\mu\beta\gamma}(\phi)-\frac{1}{2m}
C_{\mu\beta\gamma}(\phi)G^{\mu\beta\gamma}(\phi)-\frac{4}{3\sqrt{3}}\tilde{A}_{\mu\beta\gamma}G^{\mu\beta\gamma}(\phi)\right.\nn\\&+&\left.\tilde{j}_{\mu\beta\gamma}G^{\mu\beta\gamma}(\phi)\right\rbrack+
S_3[A].  \label{s3final}\eea

For simplicity we have considered only the totally symmetric source term,
$\tilde{j}_{\mu\beta\gamma}$, which will give us the correlation functions of the totally symmetric fields
$\phi_{\mu\beta\gamma}$. We have also defined the totally symmetric combination for the spin-1 field:
\be \tilde{A}_{\mu\nu\lambda}\equiv A_{\mu}\eta_{\nu\lambda}+A_{\nu}\eta_{\mu\lambda}+A_{\lambda}\eta_{\nu\mu}\ee. It is useful for the next step to notice that the
first two terms in (\ref{s3final}) are self-adjoint, i.e;
$\phi_{\mu\nu\lambda}G^{\mu\nu\lambda}(\psi)=\psi_{\mu\nu\lambda}G^{\mu\nu\lambda}(\phi)$ and
$\phi_{\mu\nu\lambda}C^{\mu\nu\lambda}(\psi)=\psi_{\mu\nu\lambda}C^{\mu\nu\lambda}(\phi)$ hold inside space-time
integrals.

Omitting the indices for simplicity, since all quantities are totally symmetric 3rd rank tensors, we propose the
following master action:

\bea S_M &=&\int \, d^3x\left \lbrack -\frac{1}{2}\phi\, G(\phi)-\frac{1}{2m}C(\phi)\,G(\phi)+\frac{1}{2m}C(\phi-\psi)\,G(\phi-\psi)-\frac{4}{3\sqrt{3}}\tilde{A}\,G(\phi)+\tilde{j}\,G(\phi)\right\rbrack \nn\\
&+& S_3[A]. \eea

\no We have used the 3rd order Chern-Simons term as the mixing term to interpolate between $SD(3)$ and $SD(4)$.
The field $\psi_{\mu\nu\lambda}$ corresponds to a new totally symmetric field. It is trivial to
observe that with the shift $\psi\to \psi+\phi$ we have the correspondence $S_M\Leftrightarrow S_{SD(3)}$. Using the
property $C(\phi)G(\sigma)=C(\sigma)G(\phi)$ one can rewrite $S_M$ as:

\be S_M =\int \,\left \lbrack -\frac{1}{2}(\phi-\sigma)\,
G(\phi-\sigma)+\frac{1}{2}\sigma\,G(\sigma)+\frac{1}{2m}C(\psi)\,G(\psi)\right\rbrack + S_3[A],\ee

\no where $\sigma$ is defined by:

\be \sigma =-\frac{C(\psi)}{m}-\frac{4}{3\sqrt{3}}\tilde{A}+\tilde{j},\label{sigma}\ee

\no making the shift $\phi \to \phi +\sigma$, as the second order term of the kind (\ref {SH}) has no particle content we end up after a
gaussian integration on $\phi$  with:

\be S_M =\int \,\left \lbrack \frac{1}{2}\sigma\,G(\sigma)+\frac{1}{2m}C(\psi)\,G(\psi)\right\rbrack +
S_3[A],\label{39}\ee

\no substituting back $(\ref{sigma})$ in $(\ref{39})$ we have, after manipulations, the complete spin-3 fourth
order self-dual model that we have found in \cite{NGEs3}:

\bea S_{SD(4)}&=&\int d^3x\s \left\lbrack \frac{1}{2m}C(\psi)G(\psi)+\frac{1}{2m^2}C(\psi)G\left\lbrack
C(\psi)\right\rbrack +\frac{4}{3\sqrt{3}m}\,C(\psi)G(\tilde{A})-\frac{1}{m}C(\psi)G(\tilde{j})\right\rbrack\nn\\
&+& S_4[A,\tilde{j}]. \label{sd4}\eea

Now the auxiliary action has gained a new second order term in derivatives, which combined with
$(\p_{\mu}A^{\mu})^2$,  is precisely the Maxwell term, written in terms of
$F_{\mu\nu}=\p_{\mu}A_{\nu}-\p_{\nu}A_{\mu}$ as:

\bea S_4[A]&=& -\frac{32}{3}\int d^3x\,\,\left\lbrack
-\frac{1}{2}F_{\mu\nu}F^{\mu\nu}+m\epsilon^{\mu\nu\alpha}A_{\mu}\p_{\nu}A_{\alpha}+m^2
A_{\mu}A^{\mu}-\tilde{j}_{\mu\beta\gamma}G^{\mu\beta\gamma}(\tilde{A})\right\rbrack. \eea

By deriving with respect to the totally symmetric source in (\ref{s3final}) and in the fourth order self-dual model
(\ref{sd4}), we have the equivalence between the correlation functions: \bea \langle
G_{\mu_1\beta_1\gamma_1}(\phi) \,...\, G_{\mu_N\beta_N\gamma_N}(\phi)\rangle_{SD3}&=&\left\langle
G_{\mu_1\beta_1\gamma_1}\left[-\frac{4C(\psi)}{3\sqrt{3}m}-\tilde{A}\right] \,...\,
G_{\mu_N\beta_N\gamma_N}\left[-\frac{4C(\psi)}{3\sqrt{3}m}-\tilde{A}\right]\right\rangle_{SD4}\nn\\
&+& C.T.\eea

\no which implies

\be
G_{\mu\beta\gamma}\left[\phi+\frac{4C(\psi)}{3\sqrt{3}m}+\tilde{A}\right]=0\ee whose general solution is pure gauge:

\be
\phi_{\mu\beta\gamma}=\p_{(\mu}\tilde{\Lambda}_{\beta\gamma)}-\frac{4C_{\mu\beta\gamma}(\psi)}{3\sqrt{3}m}+\tilde{A}_{\mu\beta\gamma}\ee
with $\tilde{\Lambda}_{\beta\gamma}$  symmetric and traceless. So we have obtained the local dual map for the
totally symmetric field $\phi_{\mu\beta\gamma}$.

\section{Final remarks and conclusion}

In our recent work \cite{NGEs3} we have obtained the fourth order self-dual model (\ref{sd4}) via NGE. In
\cite{NGEs3} we have faced the problem of not being able to find any new gauge symmetry in the fourth order term
of (\ref{sd4}) which would take us to a fifth order self-dual model, and perhaps to a sixth order
self-dual model. The authors of \cite{BHThs} have
proposed their equations of motion but not their complete actions. One could also think that, since for the
spin-1 case we have two self-dual models, and for the spin-2 four self-dual models, maybe there is a rule of the
type $2s$ where $s$ is the spin.

As an alternative method in obtaining those complete models we have tried the master action approach. As we have
seen a fundamental point when dealing with master actions consists of introducing mixing terms without particle
content. Once we can prove that the fourth order self-dual model has no particle content it would be possible to
go beyond, but unfortunately that is not the case, as we are going to analyze in what follows. The fourth order
term is given by:

\be \frac{1}{2m^2}\int d^3x\,\,C_(\psi)G\left\lbrack C(\psi)\right\rbrack = \frac{9}{2m^2} \int d^3x \s
\psi^{\mu\nu\lambda}\Box \theta_{\mu}^{ \alpha}\,\,E_{\nu}^{\s
\beta}\,\,E_{\lambda}^{\s\gamma}\psi_{\alpha\beta\gamma}. \label{4}\ee

\no In order to verify the particle content of this term we start with a lower order version of this theory with
the help of an auxiliary totally symmetric field $h_{\mu\nu\lambda}$ which is introduced in the following way:

\be S[\psi,h] = \frac{9}{m^2} \int d^3x\,\left\lbrack
h_{\mu\nu\lambda}G^{\mu\nu\lambda}(\psi)-\frac{1}{2}(h_{\mu\nu\lambda}h^{\mu\nu\lambda}-3\,\alpha
h_{\mu}h^{\mu})\right\rbrack,\label{low}\ee

\no where $h_{\mu}=\eta^{\nu\lambda}h_{\mu\nu\lambda}$. Notice that, by Gaussian integrating over
$h_{\mu\nu\lambda}$  in (\ref{low}) we have a fourth order term. In order to reproduce (\ref{4}) we set
$\alpha=7/8$. So, if and only if we have this value for $\alpha$ we have a second-order version of (\ref{4}).
This reminds us of the spin-2 case (New massive gravity \cite{bht}) where the fourth order $K$-term
$(R_{\mu\nu}R^{\mu\nu}-3R^2/8)$ can be obtained via gaussian integral over a symmetric auxiliary field
$f_{\mu\nu}$ coupled to the spin-2 Einstein tensor: \be S[g_{\mu\nu},f_{\mu\nu}]=\int d^3x\,
\sqrt{-g}\left[f_{\mu\nu}G^{\mu\nu}-\frac{m^2}{2}(f_{\mu\nu}f^{\mu\nu}-f^2)\right].\ee there is subtle
difference now, the $f^2$ term is of the Fierz-Pauli type while the $h^2$ term of (\ref{low}) does not fit in
the usual spin-3 mass term $(\alpha=1)$. Instead of integrating over $h_{\mu\nu\lambda}$, if we take the
equations of motion for $\psi_{\mu\nu\lambda}$ in (\ref{4}), we have $G_{\mu\nu\lambda}(h)=0$. We already have seen that this
immediately implies that the Ricci tensor vanishes, which on its turn implies that the Riemann tensor vanishes.
Thus, we have the general solution $h_{\mu\nu\lambda}=\p_{(\mu}\tilde{\Lambda}_{\nu\lambda)}$ with
$\eta^{\nu\lambda}\tilde{\Lambda}_{\nu\lambda}=0$. Substituting back this result in the non derivative term of
(\ref{low}), we have the rank-2 traceless theory below

\be {\cal L} = -\frac{1}{2}(h_{\mu\nu\lambda}h^{\mu\nu\lambda}-3\,\alpha h_{\mu}h^{\mu}) = \frac 32 \left\lbrack
\tilde{\Lambda}_{\mu\nu}\Box \tilde{\Lambda}^{\mu\nu} + a \, (\p^{\mu}\tilde{\Lambda}_{\mu\nu})^2 \right\rbrack
\, , \label{la} \ee

\no where we have redefined the tensors in order to get rid of the overall factor
$9/m^2$ and defined $ a = 4 \alpha - 2 $. In the specific case we are interested in, i.e., $\alpha=7/8$ we have
$a=3/2$. At this special point the theory (\ref{la}) becomes invariant under the local traceless scalar
symmetry:

\be \delta_{\Phi} \tilde{\Lambda}_{\mu\nu} = \p_{\mu}\p_{\nu} \Phi - \frac{\eta_{\mu\nu}}3 \Box \Phi  \quad .
\label{Phi} \ee

\no The equations of motion of (\ref{la}) are given by

\be \Box \tilde{\Lambda}_{\mu\nu} = \frac 34 ( \p_{\mu}A_{\nu} + \p_{\nu}A_{\mu} ) - \frac 12 \eta_{\mu\nu}
\p_{\rho}A^{\rho}  \quad , \label{eqm1} \ee

\no where we have defined the vector field

\be A^{\mu} \equiv \p_{\rho} \tilde{\Lambda}^{\rho\mu} \quad . \label{vector} \ee

\no From the equations of motion (\ref{eqm1}) it is easy to deduce the Maxwell equations:

\be \Box \, A^{\mu} - \p^{\mu} ( \p \cdot A ) = 0 \quad . \label{max} \ee

\no Due to the scalar symmetry (\ref{Phi}) one may fix the Lorentz gauge
$\p_{\mu}\p_{\nu}\tilde{\Lambda}^{\mu\nu} = \p \cdot A = 0 $. Apparently we have a massless spin-1 theory.
However, although the Lorentz gauge still has residual gauge invariance under harmonic functions $\Box \Phi = 0
$ as in the Maxwell theory, such transformations do not shift the vector field since $\delta_{\Phi}A_{\mu} =
2\Box \p_{\mu}\Phi/3 = 0 $. Consequently, we are left with $D-1=2$ degrees of freedom instead of $D-2 =  1$ as
we expect for the $3D$ Maxwell theory. The extra degree of freedom is a ghost. As a double check one can verify
that there is a double massless pole in the spin-1 sector of the propagator of the $\tilde{\Lambda}$-theory.

This is another similarity with the spin-2 case where however the fourth-order, K-term \cite{bht}, is fully
equivalent to the Maxwell theory, see \cite{bht2}. So, in the spin-2 case we have a physical massless spin-1
particle instead of a ghost. The difference comes from the nonderivative  nature of the Weyl symmetry
\cite{deserprl}   which induces a $U(1)$ change in the vector field $\delta_w h_{\mu\nu} = \eta_{\mu\nu} \, \phi
\to \delta_w A_{\mu} = \p_{\mu} \phi$ even for a harmonic function $\phi$. This allows us to get rid of the
``would be'' ghost field present in the vector field.

Anyway, in both spin-2 and spin-3 cases the nontrivial particle content of the fourth-order term invalidates its
use as a mixing term in the master action approach which avoids the transition to a possible 5th-order dual
theory. So the dualization procedure stops at the fourth-order in both cases.

In the case of the usual spin-3 mass term $\alpha=1$, $(a=2)$, the traceless model (\ref{la}) becomes exactly the W-TDIFF model in $D=3$, see \cite{blas}, which has no particle content in $D=3$. This allow us \cite{dm6} to look for a spin-3 analogue of spin-2 $NMG$ of sixth-order in derivatives.

As a final comment, we notice that in \cite{NGEs3} we have deduced the higher order spin-3 self-dual models from
the first-order one of \cite{AragoneS31} via guage embedment without any proof of spectrum equivalence which is
now clear in the master action approach used here.

\section{Acknowledgments} E.L.M and D.D thank Prof. Marc Henneaux for the reception of E.L.M at ULB, Brussels,
Belgium where part of this work was developed. The authors thank  A. Khoudeir for suggestions.  D.D thanks CNPq
(307278/2013-1) for financial support. E. L. M. thanks CNPq (449806/2014-6) and CAPES for financial support.

 \end{document}